\documentclass[12pt]{article}
\usepackage{amssymb}
\usepackage{amsmath}

\input{tcilatex}

\begin{document}

\begin{titlepage}
\title{\bf Lagrangian Mechanics on Quaternion K\"{a}hler Manifolds}
\author{ Mehmet Tekkoyun \footnote{Corresponding author. E-mail address: tekkoyun@pau.edu.tr; Tel: +902582953616; Fax: +902582953593}\\
{\small Department of Mathematics, Pamukkale University,}\\
{\small 20070 Denizli, Turkey}}
\date{\today}
\maketitle

\begin{abstract}

The aim of this study is to introduce quaterinon K\"{a}hler
analogue of Lagrangian mechanics. Finally, the geometric
and physical results related to quaternion K\"{a}hler dynamical
systems are also presented.

{\bf Keywords:} Quaternion K\"{a}hler geometry, Lagrangian
Mechanics.

{\bf PACS:} 02.40

\end{abstract}
\end{titlepage}

\section{Introduction}

It is well-known that modern differential geometry explains explicitly the
dynamics of Lagrangians. Therefore,\ we say that if $Q$ is an $m$%
-dimensional configuration manifold and $L:TQ\rightarrow \mathbf{R}$\textbf{%
\ }is a regular Lagrangian function, then there is a unique vector field $%
\xi $ on $TQ$ such that dynamics equations is given by 
\begin{equation}
i_{\xi }\Phi _{L}=dE_{L}  \label{1.1}
\end{equation}%
where $\Phi _{L}$ indicates the symplectic form. The triple $(TQ,\Phi
_{L},\xi )$ is called \textit{Lagrangian system} on the tangent bundle $TQ$ $%
.$

In literature, there are a lot of studies about Lagrangian mechanics,
formalisms, systems and equations \cite{deleon, tekkoyun} and there in.
There are real, complex, paracomplex and other analogues. It is possible to
produce different analogous in different spaces. Finding new dynamics
equations is both a new expansion and contribution to science to explain
physical events.

Quaternions were invented by Sir William Rowan Hamilton as an extension to
the complex numbers. Hamilton's defining relation is most succinctly written
as:

\begin{equation}
i^{2}=j^{2}=k^{2}=ijk=-1  \label{1.2}
\end{equation}%
If it is compared to the calculus of vectors, quaternions have slipped into
the realm of obscurity. They do however still find use in the computation of
rotations. A lot of physical laws in classical, relativistic, and quantum
mechanics can be written pleasantly by means of quaternions. Some physicists
hope they will find deeper understanding of the universe by restating basic
principles in terms of quaternion algebra. It is well-known that quaternions
are useful for representing rotations in both quantum and classical
mechanics \cite{dan} .

In this study, equations related to Lagrangian mechanical systems on
quaternion K\"{a}hler manifold have been presented.

\section{Preliminaries}

Throughout this paper, all mathematical objects and mappings are assumed to
be smooth, i.e. infinitely differentiable and Einstein convention of
summarizing is adopted. $\mathcal{F}(M)$, $\chi (M)$ and $\Lambda ^{1}(M)$
denote the set of functions on $M$, the set of vector fields on $M$ and the
set of 1-forms on $M$, respectively.

\subsection{Quaternion K\"{a}hler Manifolds}

Let $M$ be an n-dimensional manifold with a 3-dimensional vector bundle $V$
consisting of tensors of type (1,1) over $M$ satisfying condition as follows:

(a) In any coordinate neighborhood $U$ of $M$, there exists a local basis $%
\{F,G,H\}$ of $V$ such that

\begin{eqnarray}
F^{2} &=&-I,\text{ }G^{2}=-I,\text{ }H^{2}=-I  \label{2.1} \\
GH &=&-HG=F,\text{ }HF=-FH=G,\text{ }FG=-GF=H.  \notag
\end{eqnarray}%
Where $I$ denotes the identity tensor of type (1,1) in $M$. $\{F,G,H\}$ is
called a canonical local basis of the bundle $V$ in $U$. Then $V$ is called
an almost quaternion structure in $M$. The pair $(M,V)$ denotes an almost
quaternion manifold with $V$. An almost quaternion manifold $M$ is of
dimension $n=4m$ $(m\geqslant 1).$ In any almost quaternion manifold $(M,V)$%
, there is a Riemannian metric tensor field $g$ such that 
\begin{equation}
g(\phi X,Y)+g(X,\phi Y)=0  \label{2.2}
\end{equation}%
for any cross-section $\phi $ and any vector fields $X,Y$ of $M.$An almost
quaternion structure $V$ with such a Riemannian metric $g$ is called an
almost quaternion metric structure$.$ A manifold $M$ with an almost
quaternion metric structure $\{g,V\}$ is called an almost quaternion metric
manifold. The triple $(M,g,V)$ denotes an almost quaternion metric manifold.
Let $\{F,G,H\}$ be a canonical local basis of $V$ an almost quaternion
manifold $(M,g,V)$. Since each of $F,G$ and $H$ is almost Hermitian
structure with respect to $g$ metric, taking

\begin{equation}
\Phi (X,Y)=g(FX,Y),~\text{\ }\Psi (X,Y)=g(GX,Y),\text{~}\Theta (X,Y)=g(HX,Y)
\label{2.3}
\end{equation}

for any vector fields $X$ and $Y$, we see that $\Phi ,\Psi $ and $\Theta $
are local 2-forms.

Suppose that the Riemannian connection $\nabla $ of $(M,g,V)$ satisfies
conditions as follows:

$(b)$ If $\phi $ is a cross-section (local or global) of the bundle $V$,
then $V_{X}\phi $ is also a cross-section of $V$, $X$ being an arbitrary
vector field in $M$. From (\ref{2.1}) we see that the condition $(b)$ is
equivalent to condition as follows:

$(b^{\prime })$ If $F,G,H$ is a canonical local basis of $V$, then 
\begin{equation}
\nabla _{X}F=r(X)G-q(X)H,\text{ \ }\nabla _{X}G=-r(X)F+p(X)H,\text{ \ }%
\nabla _{X}H=q(X)F-p(X)G  \label{2.4}
\end{equation}

for any vector field $X$, where $p,q$ and $r$ are certain local 1-forms. If
an almost quaternion metric manifold $M$ satisfies the condition $(b)$ or $%
(b^{\prime })$, then $M$ is called a quaternion K\"{a}hler manifold and an
almost quaternion structure of $M$ is called a quaternion K\"{a}hler
structure. \cite{yano}

Let $\left\{ x_{i},x_{n+i},x_{2n+i},x_{3n+i}\right\} ,$ $i=\overline{1,n}$
be a real coordinate system on a neighborhood $U$ of $M,$ and let $\left\{ 
\frac{\partial }{\partial x_{i}},\frac{\partial }{\partial x_{n+i}},\frac{%
\partial }{\partial x_{2n+i}},\frac{\partial }{\partial x_{3n+i}}\right\} $
and $\{dx_{i},dx_{n+i},dx_{2n+i},dx_{3n+i}\}$ be natural bases over $R$ of
the tangent space $T(M)$ and the cotangent space $T^{\ast }(M)$ of $M,$
respectively$.$ Considering \cite{burdujan}, then the following expression
can be obtained%
\begin{eqnarray*}
F(\frac{\partial }{\partial x_{i}}) &=&\frac{\partial }{\partial x_{n+i}},%
\text{ }F(\frac{\partial }{\partial x_{n+i}})=-\frac{\partial }{\partial
x_{i}},\text{ }F(\frac{\partial }{\partial x_{2n+i}})=\frac{\partial }{%
\partial x_{3n+i}},\text{ }F(\frac{\partial }{\partial x_{3n+i}})=-\frac{%
\partial }{\partial x_{2n+i}} \\
G(\frac{\partial }{\partial x_{i}}) &=&\frac{\partial }{\partial x_{2n+i}},%
\text{ }G(\frac{\partial }{\partial x_{n+i}})=-\frac{\partial }{\partial
x_{3n+i}},\text{ }G(\frac{\partial }{\partial x_{2n+i}})=-\frac{\partial }{%
\partial x_{i}},\text{ }G(\frac{\partial }{\partial x_{3n+i}})=\frac{%
\partial }{\partial x_{n+i}} \\
H(\frac{\partial }{\partial x_{i}}) &=&\frac{\partial }{\partial x_{3n+i}},%
\text{ }H(\frac{\partial }{\partial x_{n+i}})=\frac{\partial }{\partial
x_{2n+i}},\text{ }H(\frac{\partial }{\partial x_{2n+i}})=-\frac{\partial }{%
\partial x_{n+i}},\text{ }H(\frac{\partial }{\partial x_{3n+i}})=-\frac{%
\partial }{\partial x_{i}}
\end{eqnarray*}

\section{Lagrangian Mechanics}

In this section, we obtain Euler-Lagrange equations for quantum and
classical mechanics by means of a canonical local basis $\{F,G,H\}$ of $V$
on quaternion K\"{a}hler manifold $(M,V).$

Firstly, let $F$ take a local basis component on the quaternion K\"{a}hler
manifold $(M,V),$ and $\left\{ x_{i},x_{n+i},x_{2n+i},x_{3n+i}\right\} $ be
its coordinate functions. Let semispray be the vector field $\xi $
determined by 
\begin{equation}
\xi =X^{i}\frac{\partial }{\partial x_{i}}+X^{n+i}\frac{\partial }{\partial
x_{n+i}}+X^{2n+i}\frac{\partial }{\partial x_{2n+i}}+X^{3n+i}\frac{\partial 
}{\partial x_{3n+i}},\,  \label{3.1}
\end{equation}%
where $X^{i}=\overset{.}{x_{i}},X^{n+i}=\overset{.}{x}_{n+i},X^{2n+i}=%
\overset{.}{x}_{2n+i},X^{3n+i}=\overset{.}{x}_{3n+i}$and the dot indicates
the derivative with respect to time $t$. The vector field defined by

\begin{equation}
V_{F}=F(\xi )=X^{i}\frac{\partial }{\partial x_{n+i}}-X^{n+i}\frac{\partial 
}{\partial x_{i}}+X^{2n+i}\frac{\partial }{\partial x_{3n+i}}-X^{3n+i}\frac{%
\partial }{\partial x_{2n+i}}  \label{3.2}
\end{equation}%
is called \textit{Liouville vector field} on the quaternion K\"{a}hler
manifold $(M,V)$. The maps given by $T,P:M\rightarrow R$ such that $T=\frac{1%
}{2}m_{i}(\overset{.}{x_{i}}^{2}+\overset{.}{x}_{n+i}^{2}+x_{2n+i}^{2}+%
\overset{.}{x}_{3n+i}^{2}),P=m_{i}gh$ are called \textit{the kinetic energy}
and \textit{the potential energy of the system,} respectively.\textit{\ }Here%
\textit{\ }$m_{i},g$ and $h$ stand for mass of a mechanical system having $m$
particles, the gravity acceleration and distance to the origin of a
mechanical system on the quaternion K\"{a}hler manifold $(M,V)$,
respectively. Then $L:M\rightarrow R$ is a map that satisfies the
conditions; i) $L=T-P$ is a \textit{Lagrangian function, ii)} the function
given by $E_{L}^{F}=V_{F}(L)-L,$ is\textit{\ energy function}.

The operator $i_{F}$ induced by $F$ and given by%
\begin{equation}
i_{F}\omega (X_{1},X_{2},...,X_{r})=\sum_{i=1}^{r}\omega
(X_{1},...,FX_{i},...,X_{r}),  \label{3.3}
\end{equation}%
is said to be \textit{vertical derivation, }where $\omega \in \wedge
^{r}{}M, $ $X_{i}\in \chi (M).$ The \textit{vertical differentiation} $d_{F}$
is defined by%
\begin{equation}
d_{F}=[i_{F},d]=i_{F}d-di_{F}  \label{3.4}
\end{equation}%
where $d$ is the usual exterior derivation. For $F$ , the closed K\"{a}hler
form is the closed 2-form given by $\Phi _{L}^{F}=-dd_{_{F}}L$ such that%
\begin{equation}
d_{_{F}}=\frac{\partial }{\partial x_{n+i}}dx_{i}-\frac{\partial }{\partial
x_{i}}dx_{n+i}+\frac{\partial }{\partial x_{3n+i}}dx_{2n+i}-\frac{\partial }{%
\partial x_{2n+i}}d_{3n+i}:\mathcal{F}(M)\rightarrow \wedge ^{1}{}M.
\label{3.5}
\end{equation}

Then%
\begin{equation}
\begin{array}{c}
\Phi _{L}^{F}=-\frac{\partial ^{2}L}{\partial x_{j}\partial x_{n+i}}%
dx_{j}\wedge dx_{i}+\frac{\partial ^{2}L}{\partial x_{j}\partial x_{i}}%
dx_{j}\wedge dx_{n+i}-\frac{\partial ^{2}L}{\partial x_{j}\partial x_{3n+i}}%
dx_{j}\wedge dx_{2n+i} \\ 
+\frac{\partial ^{2}L}{\partial x_{j}\partial x_{2n+i}}dx_{j}\wedge
dx_{3n+i}-\frac{\partial ^{2}L}{\partial x_{n+j}\partial x_{n+i}}%
dx_{n+j}\wedge dx_{i}+\frac{\partial ^{2}L}{\partial x_{n+j}\partial x_{i}}%
dx_{n+j}\wedge dx_{n+i} \\ 
-\frac{\partial ^{2}L}{\partial x_{n+j}\partial x_{3n+i}}dx_{n+j}\wedge
dx_{2n+i}+\frac{\partial ^{2}L}{\partial x_{n+j}\partial x_{2n+i}}%
dx_{n+j}\wedge dx_{3n+i}-\frac{\partial ^{2}L}{\partial x_{2n+j}\partial
x_{n+i}}dx_{2n+j}\wedge dx_{i} \\ 
+\frac{\partial ^{2}L}{\partial x_{2n+j}\partial x_{i}}dx_{2n+j}\wedge
dx_{n+i}-\frac{\partial ^{2}L}{\partial x_{2n+j}\partial x_{3n+i}}%
dx_{2n+j}\wedge dx_{2n+i}+\frac{\partial ^{2}L}{\partial x_{2n+j}\partial
x_{2n+i}}dx_{2n+j}\wedge dx_{3n+i} \\ 
-\frac{\partial ^{2}L}{\partial x_{3n+j}\partial x_{n+i}}dx_{3n+j}\wedge
dx_{i}+\frac{\partial ^{2}L}{\partial x_{3n+j}\partial x_{i}}dx_{3n+j}\wedge
dx_{n+i}-\frac{\partial ^{2}L}{\partial x_{3n+j}\partial x_{3n+i}}%
dx_{3n+j}\wedge dx_{2n+i} \\ 
+\frac{\partial ^{2}L}{\partial x_{3n+j}\partial x_{2n+i}}dx_{3n+j}\wedge
dx_{3n+i}%
\end{array}
\label{3.6}
\end{equation}%
Let $\xi $ be the second order differential equation given by \textbf{Eq. }(%
\ref{1.1}) and

\begin{equation}
\begin{array}{c}
i_{\xi }\Phi _{L}^{F}=-X^{i}\frac{\partial ^{2}L}{\partial x_{j}\partial
x_{n+i}}\delta _{i}^{j}dx_{i}+X^{i}\frac{\partial ^{2}L}{\partial
x_{j}\partial x_{n+i}}dx_{j}+X^{i}\frac{\partial ^{2}L}{\partial
x_{j}\partial x_{i}}\delta _{i}^{j}dx_{n+i} \\ 
-X^{n+i}\frac{\partial ^{2}L}{\partial x_{j}\partial x_{i}}dx_{j}-X^{i}\frac{%
\partial ^{2}L}{\partial x_{j}\partial x_{3n+i}}\delta
_{i}^{j}dx_{2n+i}+X^{2n+i}\frac{\partial ^{2}L}{\partial x_{j}\partial
x_{3n+i}}dx_{j}+X^{i}\frac{\partial ^{2}L}{\partial x_{j}\partial x_{2n+i}}%
\delta _{i}^{j}dx_{3n+i} \\ 
-X^{3n+i}\frac{\partial ^{2}L}{\partial x_{j}\partial x_{2n+i}}dx_{j}-X^{n+i}%
\frac{\partial ^{2}L}{\partial x_{n+j}\partial x_{n+i}}\delta
_{n+i}^{n+j}dx_{i}+X^{i}\frac{\partial ^{2}L}{\partial x_{n+j}\partial
x_{n+i}}dx_{n+j} \\ 
+X^{n+i}\frac{\partial ^{2}L}{\partial x_{n+j}\partial x_{i}}\delta
_{n+i}^{n+j}dx_{n+i}-X^{n+i}\frac{\partial ^{2}L}{\partial x_{n+j}\partial
x_{i}}dx_{n+j}-X^{n+i}\frac{\partial ^{2}L}{\partial x_{n+j}\partial x_{3n+i}%
}\delta _{n+i}^{n+j}dx_{2n+i} \\ 
+X^{2n+i}\frac{\partial ^{2}L}{\partial x_{n+j}\partial x_{3n+i}}%
dx_{n+j}+X^{n+i}\frac{\partial ^{2}L}{\partial x_{n+j}\partial x_{2n+i}}%
\delta _{n+i}^{n+j}dx_{3n+i}-X^{3n+i}\frac{\partial ^{2}L}{\partial
x_{n+j}\partial x_{2n+i}}dx_{n+j} \\ 
-X^{2n+i}\frac{\partial ^{2}L}{\partial x_{2n+j}\partial x_{n+i}}\delta
_{2n+i}^{2n+j}dx_{i}+X^{i}\frac{\partial ^{2}L}{\partial x_{2n+j}\partial
x_{n+i}}dx_{2n+j}+X^{2n+i}\frac{\partial ^{2}L}{\partial x_{2n+j}\partial
x_{i}}\delta _{2n+i}^{2n+j}dx_{n+i} \\ 
-X^{n+i}\frac{\partial ^{2}L}{\partial x_{2n+j}\partial x_{i}}%
dx_{2n+j}-X^{2n+i}\frac{\partial ^{2}L}{\partial x_{2n+j}\partial x_{3n+i}}%
\delta _{2n+i}^{2n+j}dx_{2n+i}+X^{2n+i}\frac{\partial ^{2}L}{\partial
x_{2n+j}\partial x_{3n+i}}dx_{2n+j} \\ 
+X^{2n+i}\frac{\partial ^{2}L}{\partial x_{2n+j}\partial x_{2n+i}}\delta
_{2n+i}^{2n+j}dx_{3n+i}-X^{3n+i}\frac{\partial ^{2}L}{\partial
x_{2n+j}\partial x_{2n+i}}dx_{2n+j}-X^{3n+i}\frac{\partial ^{2}L}{\partial
x_{3n+j}\partial x_{n+i}}\delta _{3n+i}^{3n+j}dx_{i} \\ 
+X^{i}\frac{\partial ^{2}L}{\partial x_{3n+j}\partial x_{n+i}}%
dx_{3n+j}+X^{3n+i}\frac{\partial ^{2}L}{\partial x_{3n+j}\partial x_{i}}%
\delta _{3n+i}^{3n+j}dx_{n+i}-X^{n+i}\frac{\partial ^{2}L}{\partial
x_{3n+j}\partial x_{i}}dx_{3n+j} \\ 
-X^{3n+i}\frac{\partial ^{2}L}{\partial x_{3n+j}\partial x_{3n+i}}\delta
_{3n+i}^{3n+j}dx_{2n+i}+X^{2n+i}\frac{\partial ^{2}L}{\partial
x_{3n+j}\partial x_{3n+i}}dx_{3n+j}+X^{3n+i}\frac{\partial ^{2}L}{\partial
x_{3n+j}\partial x_{2n+i}}\delta _{3n+i}^{3n+j}dx_{3n+i} \\ 
-X^{3n+i}\frac{\partial ^{2}L}{\partial x_{3n+j}\partial x_{2n+i}}dx_{3n+j}.%
\end{array}
\label{3.7}
\end{equation}

Since the closed quaternion K\"{a}hler form $\Phi _{L}^{F}$ on $(M,V)$ is
the symplectic structure, it is found

\begin{equation*}
\begin{array}{c}
E_{L}^{F}=V_{F}(L)-L=X^{i}\frac{\partial L}{\partial x_{n+i}}-X^{n+i}\frac{%
\partial L}{\partial x_{i}}+X^{2n+i}\frac{\partial L}{\partial x_{3n+i}}%
-X^{3n+i}\frac{\partial L}{\partial x_{2n+i}}-L%
\end{array}%
\end{equation*}

and hence

\begin{equation}
\begin{array}{c}
dE_{L}^{F}=X^{i}\frac{\partial ^{2}L}{\partial x_{j}\partial x_{n+i}}%
dx_{j}-X^{n+i}\frac{\partial ^{2}L}{\partial x_{j}\partial x_{i}}%
dx_{j}+X^{2n+i}\frac{\partial ^{2}L}{\partial x_{j}\partial x_{3n+i}}%
dx_{j}-X^{3n+i}\frac{\partial ^{2}L}{\partial x_{j}\partial x_{2n+i}}dx_{j}
\\ 
+X^{i}\frac{\partial ^{2}L}{\partial x_{n+j}\partial x_{n+i}}dx_{n+j}-X^{n+i}%
\frac{\partial ^{2}L}{\partial x_{n+j}\partial x_{i}}dx_{n+j}+X^{2n+i}\frac{%
\partial ^{2}L}{\partial x_{n+j}\partial x_{3n+i}}dx_{n+j}-X^{3n+i}\frac{%
\partial ^{2}L}{\partial x_{n+j}\partial x_{2n+i}}dx_{n+j} \\ 
+X^{i}\frac{\partial ^{2}L}{\partial x_{2n+j}\partial x_{n+i}}%
dx_{2n+j}-X^{n+i}\frac{\partial ^{2}L}{\partial x_{2n+j}\partial x_{i}}%
dx_{2n+j}+X^{2n+i}\frac{\partial ^{2}L}{\partial x_{2n+j}\partial x_{3n+i}}%
dx_{2n+j}-X^{3n+i}\frac{\partial ^{2}L}{\partial x_{2n+j}\partial x_{2n+i}}%
dx_{2n+j} \\ 
+X^{i}\frac{\partial ^{2}L}{\partial x_{3n+j}\partial x_{n+i}}%
dx_{3n+j}-X^{n+i}\frac{\partial ^{2}L}{\partial x_{3n+j}\partial x_{i}}%
dx_{3n+j}+X^{2n+i}\frac{\partial ^{2}L}{\partial x_{3n+j}\partial x_{3n+i}}%
dx_{3n+j}-X^{3n+i}\frac{\partial ^{2}L}{\partial x_{3n+j}\partial x_{2n+i}}%
dx_{3n+j} \\ 
-\frac{\partial L}{\partial x_{j}}dx_{j}-\frac{\partial L}{\partial x_{n+j}}%
dx_{n+j}-\frac{\partial L}{\partial x_{2n+j}}dx_{2n+j}-\frac{\partial L}{%
\partial x_{3n+j}}dx_{3n+j}%
\end{array}
\label{3.8}
\end{equation}

With the use of \textbf{Eq.} (\ref{1.1}), the following expressions can be
obtained:

\begin{equation}
\begin{array}{c}
-X^{i}\frac{\partial ^{2}L}{\partial x_{j}\partial x_{n+i}}dx_{j}+X^{i}\frac{%
\partial ^{2}L}{\partial x_{j}\partial x_{i}}dx_{n+j}-X^{i}\frac{\partial
^{2}L}{\partial x_{j}\partial x_{3n+i}}dx_{2n+j}+X^{i}\frac{\partial ^{2}L}{%
\partial x_{j}\partial x_{2n+i}}dx_{3n+j} \\ 
-X^{n+i}\frac{\partial ^{2}L}{\partial x_{n+j}\partial x_{n+i}}dx_{j}+X^{n+i}%
\frac{\partial ^{2}L}{\partial x_{n+j}\partial x_{i}}dx_{n+j}-X^{n+i}\frac{%
\partial ^{2}L}{\partial x_{n+j}\partial x_{3n+i}}dx_{2n+j} \\ 
+X^{n+i}\frac{\partial ^{2}L}{\partial x_{n+j}\partial x_{2n+i}}%
dx_{3n+j}-X^{2n+i}\frac{\partial ^{2}L}{\partial x_{2n+j}\partial x_{n+i}}%
dx_{j}+X^{2n+i}\frac{\partial ^{2}L}{\partial x_{2n+j}\partial x_{i}}dx_{n+j}
\\ 
-X^{2n+i}\frac{\partial ^{2}L}{\partial x_{2n+j}\partial x_{3n+i}}%
dx_{2n+j}+X^{2n+i}\frac{\partial ^{2}L}{\partial x_{2n+j}\partial x_{2n+i}}%
dx_{3n+j}-X^{3n+i}\frac{\partial ^{2}L}{\partial x_{3n+j}\partial x_{n+i}}%
dx_{j} \\ 
+X^{3n+i}\frac{\partial ^{2}L}{\partial x_{3n+j}\partial x_{i}}%
dx_{n+j}-X^{3n+i}\frac{\partial ^{2}L}{\partial x_{3n+j}\partial x_{3n+i}}%
dx_{2n+j}+X^{3n+i}\frac{\partial ^{2}L}{\partial x_{3n+j}\partial x_{2n+i}}%
dx_{3n+j} \\ 
+\frac{\partial L}{\partial x_{j}}dx_{j}+\frac{\partial L}{\partial x_{n+j}}%
dx_{n+j}+\frac{\partial L}{\partial x_{2n+j}}dx_{2n+j}+\frac{\partial L}{%
\partial x_{3n+j}}dx_{3n+j}=0%
\end{array}
\label{3.9}
\end{equation}%
If a curve denoted by $\alpha :R\rightarrow M$ is considered to be an
integral curve of $\xi ,$ then we calculate the following equation: 
\begin{equation}
\begin{array}{c}
-X^{i}\frac{\partial ^{2}L}{\partial x_{j}\partial x_{n+i}}dx_{j}-X^{n+i}%
\frac{\partial ^{2}L}{\partial x_{n+j}\partial x_{n+i}}dx_{j}-X^{2n+i}\frac{%
\partial ^{2}L}{\partial x_{2n+j}\partial x_{n+i}}dx_{j}-X^{3n+i}\frac{%
\partial ^{2}L}{\partial x_{3n+j}\partial x_{n+i}}dx_{j} \\ 
+X^{i}\frac{\partial ^{2}L}{\partial x_{j}\partial x_{i}}dx_{n+j}+X^{n+i}%
\frac{\partial ^{2}L}{\partial x_{n+j}\partial x_{i}}dx_{n+j}+X^{2n+i}\frac{%
\partial ^{2}L}{\partial x_{2n+j}\partial x_{i}}dx_{n+j}+X^{3n+i}\frac{%
\partial ^{2}L}{\partial x_{3n+j}\partial x_{i}}dx_{n+j} \\ 
-X^{i}\frac{\partial ^{2}L}{\partial x_{j}\partial x_{3n+i}}dx_{2n+j}-X^{n+i}%
\frac{\partial ^{2}L}{\partial x_{n+j}\partial x_{3n+i}}dx_{2n+j}-X^{2n+i}%
\frac{\partial ^{2}L}{\partial x_{2n+j}\partial x_{3n+i}}dx_{2n+j}-X^{3n+i}%
\frac{\partial ^{2}L}{\partial x_{3n+j}\partial x_{3n+i}}dx_{2n+j} \\ 
+X^{i}\frac{\partial ^{2}L}{\partial x_{j}\partial x_{2n+i}}dx_{3n+j}+X^{n+i}%
\frac{\partial ^{2}L}{\partial x_{n+j}\partial x_{2n+i}}dx_{3n+j}+X^{2n+i}%
\frac{\partial ^{2}L}{\partial x_{2n+j}\partial x_{2n+i}}dx_{3n+j}+X^{3n+i}%
\frac{\partial ^{2}L}{\partial x_{3n+j}\partial x_{2n+i}}dx_{3n+j} \\ 
+\frac{\partial L}{\partial x_{j}}dx_{j}+\frac{\partial L}{\partial x_{n+j}}%
dx_{n+j}+\frac{\partial L}{\partial x_{2n+j}}dx_{2n+j}+\frac{\partial L}{%
\partial x_{3n+j}}dx_{3n+j}=0,%
\end{array}
\label{3.10}
\end{equation}%
alternatively 
\begin{equation}
\begin{array}{c}
-[X^{i}\frac{\partial ^{2}L}{\partial x_{j}\partial x_{n+i}}+X^{n+i}\frac{%
\partial ^{2}L}{\partial x_{n+j}\partial x_{n+i}}+X^{2n+i}\frac{\partial
^{2}L}{\partial x_{2n+j}\partial x_{n+i}}+X^{3n+i}\frac{\partial ^{2}L}{%
\partial x_{3n+j}\partial x_{n+i}}]dx_{j}+\frac{\partial L}{\partial x_{j}}%
dx_{j} \\ 
+[X^{i}\frac{\partial ^{2}L}{\partial x_{j}\partial x_{i}}+X^{n+i}\frac{%
\partial ^{2}L}{\partial x_{n+j}\partial x_{i}}+X^{2n+i}\frac{\partial ^{2}L%
}{\partial x_{2n+j}\partial x_{i}}+X^{3n+i}\frac{\partial ^{2}L}{\partial
x_{3n+j}\partial x_{i}}]dx_{n+j}+\frac{\partial L}{\partial x_{n+j}}dx_{n+j}
\\ 
-[X^{i}\frac{\partial ^{2}L}{\partial x_{j}\partial x_{3n+i}}+X^{n+i}\frac{%
\partial ^{2}L}{\partial x_{n+j}\partial x_{3n+i}}+X^{2n+i}\frac{\partial
^{2}L}{\partial x_{2n+j}\partial x_{3n+i}}+X^{3n+i}\frac{\partial ^{2}L}{%
\partial x_{3n+j}\partial x_{3n+i}}]dx_{2n+j}+\frac{\partial L}{\partial
x_{2n+j}}dx_{2n+j} \\ 
+[X^{i}\frac{\partial ^{2}L}{\partial x_{j}\partial x_{2n+i}}+X^{n+i}\frac{%
\partial ^{2}L}{\partial x_{n+j}\partial x_{2n+i}}+X^{2n+i}\frac{\partial
^{2}L}{\partial x_{2n+j}\partial x_{2n+i}}+X^{3n+i}\frac{\partial ^{2}L}{%
\partial x_{3n+j}\partial x_{2n+i}}]dx_{3n+j}+\frac{\partial L}{\partial
x_{3n+j}}dx_{3n+j}=0.%
\end{array}
\label{3.11}
\end{equation}

Then we obtain the equations 
\begin{equation}
\begin{array}{l}
\frac{\partial }{\partial t}\left( \frac{\partial L}{\partial x_{i}}\right) +%
\frac{\partial L}{\partial x_{n+i}}=0,\text{ }\frac{\partial }{\partial t}%
\left( \frac{\partial L}{\partial x_{n+i}}\right) -\frac{\partial L}{%
\partial x_{i}}=0, \\ 
\,\frac{\partial }{\partial t}\left( \frac{\partial L}{\partial x_{2n+i}}%
\right) +\frac{\partial L}{\partial x_{3n+i}}=0,\,\text{\ }\frac{\partial }{%
\partial t}\left( \frac{\partial L}{\partial x_{3n+i}}\right) -\frac{%
\partial L}{\partial x_{2n+i}}=0,%
\end{array}
\label{3.12}
\end{equation}%
such that the equations obtained in \textbf{Eq. }(\ref{3.12}) are said to be 
\textit{Euler-Lagrange equations} structured on quaternion K\"{a}hler
manifold $(M,V)$ by means of $\Phi _{L}^{F}$ and thus the triple $(M,\Phi
_{L}^{F},\xi )$ is said to be a \textit{mechanical system }on quaternion K%
\"{a}hler manifold $(M,V)$\textit{.}

Secondly, we find Euler-Lagrange equations for quantum and classical
mechanics by means of $\Phi _{L}^{G}$ on quaternion K\"{a}hler manifold $%
(M,V).$

Consider $G$ be another local basis component on the quaternion K\"{a}hler
manifold $(M,V).$ Let $\xi $ take as in \textbf{Eq.} (\ref{3.1}). In the
case, the vector field given by

\begin{equation}
V_{G}=G(\xi )=X^{i}\frac{\partial }{\partial x_{2n+i}}-X^{n+i}\frac{\partial 
}{\partial x_{3n+i}}-X^{2n+i}\frac{\partial }{\partial x_{i}}+X^{3n+i}\frac{%
\partial }{\partial x_{n+i}}  \label{3.13}
\end{equation}%
is \textit{Liouville vector field} on the quaternion K\"{a}hler manifold $%
(M,V)$. The function given by $E_{L}^{G}=V_{G}(L)-L$ is\textit{\ energy
function}. Then the operator $i_{G}$ induced by $G$ and denoted by%
\begin{equation}
i_{G}\omega (X_{1},X_{2},...,X_{r})=\sum_{i=1}^{r}\omega
(X_{1},...,GX_{i},...,X_{r})  \label{3.14}
\end{equation}%
is \textit{vertical derivation, }where $\omega \in \wedge ^{r}{}M,$ $%
X_{i}\in \chi (M).$ The \textit{vertical differentiation} $d_{G}$ are
defined by%
\begin{equation}
d_{G}=[i_{G},d]=i_{G}d-di_{G}  \label{3.15}
\end{equation}%
where $d$ is the usual exterior derivation. Since taking into considering $%
G, $ the closed K\"{a}hler form is the closed 2-form given by $\Phi
_{L}^{G}=-dd_{_{G}}L$ such that%
\begin{equation}
d_{_{G}}=\frac{\partial }{\partial x_{2n+i}}dx_{i}-\frac{\partial }{\partial
x_{3n+i}}dx_{n+i}-\frac{\partial }{\partial x_{i}}dx_{2n+i}+\frac{\partial }{%
\partial x_{n+i}}d_{3n+i}:\mathcal{F}(M)\rightarrow \wedge ^{1}{}M.
\label{3.16}
\end{equation}

Then we have%
\begin{equation}
\begin{array}{c}
\Phi _{L}^{G}=-\frac{\partial ^{2}L}{\partial x_{j}\partial x_{2n+i}}%
dx_{j}\wedge dx_{i}+\frac{\partial ^{2}L}{\partial x_{j}\partial x_{3n+i}}%
dx_{j}\wedge dx_{n+i}+\frac{\partial ^{2}L}{\partial x_{j}\partial x_{i}}%
dx_{j}\wedge dx_{2n+i} \\ 
-\frac{\partial ^{2}L}{\partial x_{j}\partial x_{n+i}}dx_{j}\wedge dx_{3n+i}-%
\frac{\partial ^{2}L}{\partial x_{n+j}\partial x_{2n+i}}dx_{n+j}\wedge
dx_{i}+\frac{\partial ^{2}L}{\partial x_{n+j}\partial x_{3n+i}}%
dx_{n+j}\wedge dx_{n+i} \\ 
+\frac{\partial ^{2}L}{\partial x_{n+j}\partial x_{i}}dx_{n+j}\wedge
dx_{2n+i}-\frac{\partial ^{2}L}{\partial x_{n+j}\partial x_{n+i}}%
dx_{n+j}\wedge dx_{3n+i}-\frac{\partial ^{2}L}{\partial x_{2n+j}\partial
x_{2n+i}}dx_{2n+j}\wedge dx_{i} \\ 
+\frac{\partial ^{2}L}{\partial x_{2n+j}\partial x_{3n+i}}dx_{2n+j}\wedge
dx_{n+i}+\frac{\partial ^{2}L}{\partial x_{2n+j}\partial x_{i}}%
dx_{2n+j}\wedge dx_{2n+i}-\frac{\partial ^{2}L}{\partial x_{2n+j}\partial
x_{n+i}}dx_{2n+j}\wedge dx_{3n+i} \\ 
-\frac{\partial ^{2}L}{\partial x_{3n+j}\partial x_{2n+i}}dx_{3n+j}\wedge
dx_{i}+\frac{\partial ^{2}L}{\partial x_{3n+j}\partial x_{3n+i}}%
dx_{3n+j}\wedge dx_{n+i}+\frac{\partial ^{2}L}{\partial x_{3n+j}\partial
x_{i}}dx_{3n+j}\wedge dx_{2n+i} \\ 
-\frac{\partial ^{2}L}{\partial x_{3n+j}\partial x_{n+i}}dx_{3n+j}\wedge
dx_{3n+i}.%
\end{array}
\label{3.17}
\end{equation}%
Let $\xi $ be differential equation yielding \textbf{Eq. }(\ref{1.1}) and,
then it follows

\begin{equation}
\begin{array}{c}
i_{\xi }\Phi _{L}^{G}=-X^{i}\frac{\partial ^{2}L}{\partial x_{j}\partial
x_{2n+i}}\delta _{i}^{j}dx_{i}+X^{i}\frac{\partial ^{2}L}{\partial
x_{j}\partial x_{2n+i}}dx_{j}+X^{i}\frac{\partial ^{2}L}{\partial
x_{j}\partial x_{3n+i}}\delta _{i}^{j}dx_{n+i} \\ 
-X^{n+i}\frac{\partial ^{2}L}{\partial x_{j}\partial x_{3n+i}}dx_{j}+X^{i}%
\frac{\partial ^{2}L}{\partial x_{j}\partial x_{i}}\delta
_{i}^{j}dx_{2n+i}-X^{2n+i}\frac{\partial ^{2}L}{\partial x_{j}\partial x_{i}}%
dx_{j}-X^{i}\frac{\partial ^{2}L}{\partial x_{j}\partial x_{n+i}}\delta
_{i}^{j}dx_{3n+i} \\ 
+X^{3n+i}\frac{\partial ^{2}L}{\partial x_{j}\partial x_{n+i}}dx_{j}-X^{n+i}%
\frac{\partial ^{2}L}{\partial x_{n+j}\partial x_{2n+i}}\delta
_{n+i}^{n+j}dx_{i}+X^{i}\frac{\partial ^{2}L}{\partial x_{n+j}\partial
x_{2n+i}}dx_{n+j} \\ 
+X^{n+i}\frac{\partial ^{2}L}{\partial x_{n+j}\partial x_{3n+i}}\delta
_{n+i}^{n+j}dx_{n+i}-X^{n+i}\frac{\partial ^{2}L}{\partial x_{n+j}\partial
x_{3n+i}}dx_{n+j}+X^{n+i}\frac{\partial ^{2}L}{\partial x_{n+j}\partial x_{i}%
}\delta _{n+i}^{n+j}dx_{2n+i} \\ 
-X^{2n+i}\frac{\partial ^{2}L}{\partial x_{n+j}\partial x_{i}}%
dx_{n+j}-X^{n+i}\frac{\partial ^{2}L}{\partial x_{n+j}\partial x_{n+i}}%
\delta _{n+i}^{n+j}dx_{3n+i}+X^{3n+i}\frac{\partial ^{2}L}{\partial
x_{n+j}\partial x_{n+i}}dx_{n+j} \\ 
-X^{2n+i}\frac{\partial ^{2}L}{\partial x_{2n+j}\partial x_{2n+i}}\delta
_{2n+i}^{2n+j}dx_{i}+X^{i}\frac{\partial ^{2}L}{\partial x_{2n+j}\partial
x_{2n+i}}dx_{2n+j}+X^{2n+i}\frac{\partial ^{2}L}{\partial x_{2n+j}\partial
x_{3n+i}}\delta _{2n+i}^{2n+j}dx_{n+i} \\ 
-X^{n+i}\frac{\partial ^{2}L}{\partial x_{2n+j}\partial x_{3n+i}}%
dx_{2n+j}+X^{2n+i}\frac{\partial ^{2}L}{\partial x_{2n+j}\partial x_{i}}%
\delta _{2n+i}^{2n+j}dx_{2n+i}-X^{2n+i}\frac{\partial ^{2}L}{\partial
x_{2n+j}\partial x_{i}}dx_{2n+j} \\ 
-X^{2n+i}\frac{\partial ^{2}L}{\partial x_{2n+j}\partial x_{n+i}}\delta
_{2n+i}^{2n+j}dx_{3n+i}+X^{3n+i}\frac{\partial ^{2}L}{\partial
x_{2n+j}\partial x_{n+i}}dx_{2n+j}-X^{3n+i}\frac{\partial ^{2}L}{\partial
x_{3n+j}\partial x_{2n+i}}\delta _{3n+i}^{3n+j}dx_{i} \\ 
+X^{i}\frac{\partial ^{2}L}{\partial x_{3n+j}\partial x_{2n+i}}%
dx_{3n+j}+X^{3n+i}\frac{\partial ^{2}L}{\partial x_{3n+j}\partial x_{3n+i}}%
\delta _{3n+i}^{3n+j}dx_{n+i}-X^{n+i}\frac{\partial ^{2}L}{\partial
x_{3n+j}\partial x_{3n+i}}dx_{3n+j} \\ 
+X^{3n+i}\frac{\partial ^{2}L}{\partial x_{3n+j}\partial x_{i}}\delta
_{3n+i}^{3n+j}dx_{2n+i}-X^{2n+i}\frac{\partial ^{2}L}{\partial
x_{3n+j}\partial x_{i}}dx_{3n+j}-X^{3n+i}\frac{\partial ^{2}L}{\partial
x_{3n+j}\partial x_{n+i}}\delta _{3n+i}^{3n+j}dx_{3n+i} \\ 
+X^{3n+i}\frac{\partial ^{2}L}{\partial x_{3n+j}\partial x_{n+i}}dx_{3n+j}%
\end{array}
\label{3.18}
\end{equation}

Since the closed K\"{a}hler form $\Phi _{L}^{G}$ on $M$ is the symplectic
structure, it is gotten

\begin{equation}
E_{L}^{G}=V_{G}(L)-L=X^{i}\frac{\partial L}{\partial x_{2n+i}}-X^{n+i}\frac{%
\partial L}{\partial x_{3n+i}}-X^{2n+i}\frac{\partial L}{\partial x_{i}}%
+X^{3n+i}\frac{\partial L}{\partial x_{n+i}}-L  \label{3.19}
\end{equation}

and hence

\begin{equation}
\begin{array}{c}
dE_{L}^{G}=X^{i}\frac{\partial ^{2}L}{\partial x_{j}\partial x_{2n+i}}%
dx_{j}-X^{n+i}\frac{\partial ^{2}L}{\partial x_{j}\partial x_{3n+i}}%
dx_{j}-X^{2n+i}\frac{\partial ^{2}L}{\partial x_{j}\partial x_{i}}%
dx_{j}+X^{3n+i}\frac{\partial ^{2}L}{\partial x_{j}\partial x_{n+i}}dx_{j}
\\ 
+X^{i}\frac{\partial ^{2}L}{\partial x_{n+j}\partial x_{2n+i}}%
dx_{n+j}-X^{n+i}\frac{\partial ^{2}L}{\partial x_{n+j}\partial x_{3n+i}}%
dx_{n+j}-X^{2n+i}\frac{\partial ^{2}L}{\partial x_{n+j}\partial x_{i}}%
dx_{n+j}+X^{3n+i}\frac{\partial ^{2}L}{\partial x_{n+j}\partial x_{n+i}}%
dx_{n+j} \\ 
+X^{i}\frac{\partial ^{2}L}{\partial x_{2n+j}\partial x_{2n+i}}%
dx_{2n+j}-X^{n+i}\frac{\partial ^{2}L}{\partial x_{2n+j}\partial x_{3n+i}}%
dx_{2n+j}-X^{2n+i}\frac{\partial ^{2}L}{\partial x_{2n+j}\partial x_{i}}%
dx_{2n+j}+X^{3n+i}\frac{\partial ^{2}L}{\partial x_{2n+j}\partial x_{n+i}}%
dx_{2n+j} \\ 
+X^{i}\frac{\partial ^{2}L}{\partial x_{3n+j}\partial x_{2n+i}}%
dx_{3n+j}-X^{n+i}\frac{\partial ^{2}L}{\partial x_{3n+j}\partial x_{3n+i}}%
dx_{3n+j}-X^{2n+i}\frac{\partial ^{2}L}{\partial x_{3n+j}\partial x_{i}}%
dx_{3n+j}+X^{3n+i}\frac{\partial ^{2}L}{\partial x_{3n+j}\partial x_{n+i}}%
dx_{3n+j} \\ 
-\frac{\partial L}{\partial x_{j}}dx_{j}-\frac{\partial L}{\partial x_{n+j}}%
dx_{n+j}-\frac{\partial L}{\partial x_{2n+j}}dx_{2n+j}-\frac{\partial L}{%
\partial x_{3n+j}}dx_{3n+j}%
\end{array}
\label{3.20}
\end{equation}

By means of \textbf{Eq.} (\ref{1.1}), we calculate

\begin{equation}
\begin{array}{c}
-X^{i}\frac{\partial ^{2}L}{\partial x_{j}\partial x_{2n+i}}dx_{j}+X^{i}%
\frac{\partial ^{2}L}{\partial x_{j}\partial x_{3n+i}}dx_{n+j}+X^{i}\frac{%
\partial ^{2}L}{\partial x_{j}\partial x_{i}}dx_{2n+j}-X^{i}\frac{\partial
^{2}L}{\partial x_{j}\partial x_{n+i}}dx_{3n+j} \\ 
-X^{n+i}\frac{\partial ^{2}L}{\partial x_{n+j}\partial x_{2n+i}}%
dx_{j}+X^{n+i}\frac{\partial ^{2}L}{\partial x_{n+j}\partial x_{3n+i}}%
dx_{n+j}+X^{n+i}\frac{\partial ^{2}L}{\partial x_{n+j}\partial x_{i}}%
dx_{2n+j} \\ 
-X^{n+i}\frac{\partial ^{2}L}{\partial x_{n+j}\partial x_{n+i}}%
dx_{3n+j}-X^{2n+i}\frac{\partial ^{2}L}{\partial x_{2n+j}\partial x_{2n+i}}%
dx_{j}+X^{2n+i}\frac{\partial ^{2}L}{\partial x_{2n+j}\partial x_{3n+i}}%
dx_{n+j} \\ 
+X^{2n+i}\frac{\partial ^{2}L}{\partial x_{2n+j}\partial x_{i}}%
dx_{2n+j}-X^{2n+i}\frac{\partial ^{2}L}{\partial x_{2n+j}\partial x_{n+i}}%
dx_{3n+j}-X^{3n+i}\frac{\partial ^{2}L}{\partial x_{3n+j}\partial x_{2n+i}}%
dx_{j} \\ 
+X^{3n+i}\frac{\partial ^{2}L}{\partial x_{3n+j}\partial x_{3n+i}}%
dx_{n+j}+X^{3n+i}\frac{\partial ^{2}L}{\partial x_{3n+j}\partial x_{i}}%
dx_{2n+j}-X^{3n+i}\frac{\partial ^{2}L}{\partial x_{3n+j}\partial x_{n+i}}%
dx_{3n+j} \\ 
+\frac{\partial L}{\partial x_{j}}dx_{j}+\frac{\partial L}{\partial x_{n+j}}%
dx_{n+j}+\frac{\partial L}{\partial x_{2n+j}}dx_{2n+j}+\frac{\partial L}{%
\partial x_{3n+j}}dx_{3n+j}=0%
\end{array}
\label{3.21}
\end{equation}%
If a curve, defined by $\alpha :R\rightarrow M,$ is an integral curve of $%
\xi ,$ then we obtain 
\begin{equation}
\begin{array}{c}
-X^{i}\frac{\partial ^{2}L}{\partial x_{j}\partial x_{2n+i}}dx_{j}-X^{n+i}%
\frac{\partial ^{2}L}{\partial x_{n+j}\partial x_{2n+i}}dx_{j}-X^{2n+i}\frac{%
\partial ^{2}L}{\partial x_{2n+j}\partial x_{2n+i}}dx_{j}-X^{3n+i}\frac{%
\partial ^{2}L}{\partial x_{3n+j}\partial x_{2n+i}}dx_{j} \\ 
+X^{i}\frac{\partial ^{2}L}{\partial x_{j}\partial x_{3n+i}}dx_{n+j}+X^{n+i}%
\frac{\partial ^{2}L}{\partial x_{n+j}\partial x_{3n+i}}dx_{n+j}+X^{2n+i}%
\frac{\partial ^{2}L}{\partial x_{2n+j}\partial x_{3n+i}}dx_{n+j}+X^{3n+i}%
\frac{\partial ^{2}L}{\partial x_{3n+j}\partial x_{3n+i}}dx_{n+j} \\ 
+X^{i}\frac{\partial ^{2}L}{\partial x_{j}\partial x_{i}}dx_{2n+j}+X^{n+i}%
\frac{\partial ^{2}L}{\partial x_{n+j}\partial x_{i}}dx_{2n+j}+X^{2n+i}\frac{%
\partial ^{2}L}{\partial x_{2n+j}\partial x_{i}}dx_{2n+j}+X^{3n+i}\frac{%
\partial ^{2}L}{\partial x_{3n+j}\partial x_{i}}dx_{2n+j} \\ 
-X^{i}\frac{\partial ^{2}L}{\partial x_{j}\partial x_{n+i}}dx_{3n+j}-X^{n+i}%
\frac{\partial ^{2}L}{\partial x_{n+j}\partial x_{n+i}}dx_{3n+j}-X^{2n+i}%
\frac{\partial ^{2}L}{\partial x_{2n+j}\partial x_{n+i}}dx_{3n+j}-X^{3n+i}%
\frac{\partial ^{2}L}{\partial x_{3n+j}\partial x_{n+i}}dx_{3n+j} \\ 
+\frac{\partial L}{\partial x_{j}}dx_{j}+\frac{\partial L}{\partial x_{n+j}}%
dx_{n+j}+\frac{\partial L}{\partial x_{2n+j}}dx_{2n+j}+\frac{\partial L}{%
\partial x_{3n+j}}dx_{3n+j}=0%
\end{array}
\label{3.22}
\end{equation}%
or%
\begin{equation}
\begin{array}{c}
-[X^{i}\frac{\partial ^{2}L}{\partial x_{j}\partial x_{2n+i}}+X^{n+i}\frac{%
\partial ^{2}L}{\partial x_{n+j}\partial x_{2n+i}}+X^{2n+i}\frac{\partial
^{2}L}{\partial x_{2n+j}\partial x_{2n+i}}+X^{3n+i}\frac{\partial ^{2}L}{%
\partial x_{3n+j}\partial x_{2n+i}}]dx_{j}+\frac{\partial L}{\partial x_{j}}%
dx_{j} \\ 
+[X^{i}\frac{\partial ^{2}L}{\partial x_{j}\partial x_{3n+i}}+X^{n+i}\frac{%
\partial ^{2}L}{\partial x_{n+j}\partial x_{3n+i}}+X^{2n+i}\frac{\partial
^{2}L}{\partial x_{2n+j}\partial x_{3n+i}}+X^{3n+i}\frac{\partial ^{2}L}{%
\partial x_{3n+j}\partial x_{3n+i}}]dx_{n+j}+\frac{\partial L}{\partial
x_{n+j}}dx_{n+j} \\ 
+[X^{i}\frac{\partial ^{2}L}{\partial x_{j}\partial x_{i}}+X^{n+i}\frac{%
\partial ^{2}L}{\partial x_{n+j}\partial x_{i}}+X^{2n+i}\frac{\partial ^{2}L%
}{\partial x_{2n+j}\partial x_{i}}+X^{3n+i}\frac{\partial ^{2}L}{\partial
x_{3n+j}\partial x_{i}}]dx_{2n+j}+\frac{\partial L}{\partial x_{2n+j}}%
dx_{2n+j} \\ 
-[X^{i}\frac{\partial ^{2}L}{\partial x_{j}\partial x_{n+i}}+X^{n+i}\frac{%
\partial ^{2}L}{\partial x_{n+j}\partial x_{n+i}}+X^{2n+i}\frac{\partial
^{2}L}{\partial x_{2n+j}\partial x_{n+i}}+X^{3n+i}\frac{\partial ^{2}L}{%
\partial x_{3n+j}\partial x_{n+i}}]dx_{3n+j}+\frac{\partial L}{\partial
x_{3n+j}}dx_{3n+j}=0%
\end{array}
\label{3.23}
\end{equation}%
Then the equations are found: 
\begin{equation}
\begin{array}{l}
\,\frac{\partial }{\partial t}\left( \frac{\partial L}{\partial x_{i}}%
\right) +\frac{\partial L}{\partial x_{2n+i}}=0,\frac{\partial }{\partial t}%
\left( \frac{\partial L}{\partial x_{n+i}}\right) -\frac{\partial L}{%
\partial x_{3n+i}}=0, \\ 
\text{ }\frac{\partial }{\partial t}\left( \frac{\partial L}{\partial
x_{2n+i}}\right) -\frac{\partial L}{\partial x_{i}}=0,\,\,\,\,\frac{\partial 
}{\partial t}\left( \frac{\partial L}{\partial x_{3n+i}}\right) +\frac{%
\partial L}{\partial x_{n+i}}=0.\,%
\end{array}
\label{3.24}
\end{equation}%
Thus the equations obtained in \textbf{Eq. }(\ref{3.24}) are called \textit{%
Euler-Lagrange equations} structured by means of $\Phi _{L}^{G}$ on
quaternion K\"{a}hler manifold $(M,V)$ and thus the triple $(M,\Phi
_{L}^{G},\xi )$ can be called to be a \textit{mechanical system }on
quaternion K\"{a}hler manifold $(M,V)$\textit{.}

Thirdly, we introduce Euler-Lagrange equations for quantum and classical
mechanics by means of $\Phi _{L}^{H}$ on quaternion K\"{a}hler manifold $%
(M,V).$

Let $H$ be a local basis on the quaternion K\"{a}hler manifold $(M,V).$Let
semispray $\xi $ give as in \textbf{Eq.}(\ref{3.1}). Therefore, \textit{%
Liouville vector field} on the quaternion K\"{a}hler manifold $(M,V)$ is the
vector field given by

\begin{equation}
V_{H}=H(\xi )=X^{i}\frac{\partial }{\partial x_{3n+i}}+X^{n+i}\frac{\partial 
}{\partial x_{2n+i}}-X^{2n+i}\frac{\partial }{\partial x_{n+i}}-X^{3n+i}%
\frac{\partial }{\partial x_{i}}.  \label{3.25}
\end{equation}%
The function given by $E_{L}^{H}=V_{H}(L)-L$ is\textit{\ energy function}.
The function $i_{H}$ induced by $H$ and shown by%
\begin{equation}
i_{H}\omega (X_{1},X_{2},...,X_{r})=\sum_{i=1}^{r}\omega
(X_{1},...,HX_{i},...,X_{r}),  \label{3.26}
\end{equation}%
is said to be \textit{vertical derivation, }where $\omega \in \wedge
^{r}{}M, $ $X_{i}\in \chi (M).$ The \textit{vertical differentiation} $d_{H}$
is denoted by%
\begin{equation}
d_{H}=[i_{H},d]=i_{H}d-di_{H},  \label{3.27}
\end{equation}%
where $d$ is the usual exterior derivation. Considering $H$ , the closed K%
\"{a}hler form is the closed 2-form given by $\Phi _{L}^{H}=-dd_{_{H}}L$
such that%
\begin{equation}
d_{_{H}}=\frac{\partial }{\partial x_{3n+i}}dx_{i}+\frac{\partial }{\partial
x_{2n+i}}dx_{n+i}-\frac{\partial }{\partial x_{n+i}}dx_{2n+i}-\frac{\partial 
}{\partial x_{i}}d_{3n+i}:\mathcal{F}(M)\rightarrow \wedge ^{1}{}M
\label{3.28}
\end{equation}

Then we get%
\begin{equation}
\begin{array}{c}
\Phi _{L}^{H}=-\frac{\partial ^{2}L}{\partial x_{j}\partial x_{3n+i}}%
dx_{j}\wedge dx_{i}-\frac{\partial ^{2}L}{\partial x_{j}\partial x_{2n+i}}%
dx_{j}\wedge dx_{n+i}+\frac{\partial ^{2}L}{\partial x_{j}\partial x_{n+i}}%
dx_{j}\wedge dx_{2n+i} \\ 
+\frac{\partial ^{2}L}{\partial x_{j}\partial x_{i}}dx_{j}\wedge dx_{3n+i}-%
\frac{\partial ^{2}L}{\partial x_{n+j}\partial x_{3n+i}}dx_{n+j}\wedge
dx_{i}-\frac{\partial ^{2}L}{\partial x_{n+j}\partial x_{2n+i}}%
dx_{n+j}\wedge dx_{n+i} \\ 
+\frac{\partial ^{2}L}{\partial x_{n+j}\partial x_{n+i}}dx_{n+j}\wedge
dx_{2n+i}+\frac{\partial ^{2}L}{\partial x_{n+j}\partial x_{i}}%
dx_{n+j}\wedge dx_{3n+i}-\frac{\partial ^{2}L}{\partial x_{2n+j}\partial
x_{3n+i}}dx_{2n+j}\wedge dx_{i} \\ 
-\frac{\partial ^{2}L}{\partial x_{2n+j}\partial x_{2n+i}}dx_{2n+j}\wedge
dx_{n+i}+\frac{\partial ^{2}L}{\partial x_{2n+j}\partial x_{n+i}}%
dx_{2n+j}\wedge dx_{2n+i}+\frac{\partial ^{2}L}{\partial x_{2n+j}\partial
x_{i}}dx_{2n+j}\wedge dx_{3n+i} \\ 
-\frac{\partial ^{2}L}{\partial x_{3n+j}\partial x_{3n+i}}dx_{3n+j}\wedge
dx_{i}-\frac{\partial ^{2}L}{\partial x_{3n+j}\partial x_{2n+i}}%
dx_{3n+j}\wedge dx_{n+i}+\frac{\partial ^{2}L}{\partial x_{3n+j}\partial
x_{n+i}}dx_{3n+j}\wedge dx_{2n+i} \\ 
+\frac{\partial ^{2}L}{\partial x_{3n+j}\partial x_{i}}dx_{3n+j}\wedge
dx_{3n+i}%
\end{array}
\label{3.29}
\end{equation}%
Let $\xi $ be the semispray given by \textbf{Eq. }(\ref{1.1}) and, then we
find

\begin{equation}
\begin{array}{c}
i_{\xi }\Phi _{L}^{H}=-X^{i}\frac{\partial ^{2}L}{\partial x_{j}\partial
x_{3n+i}}\delta _{i}^{j}dx_{i}+X^{i}\frac{\partial ^{2}L}{\partial
x_{j}\partial x_{3n+i}}dx_{j}-X^{i}\frac{\partial ^{2}L}{\partial
x_{j}\partial x_{2n+i}}\delta _{i}^{j}dx_{n+i} \\ 
+X^{n+i}\frac{\partial ^{2}L}{\partial x_{j}\partial x_{2n+i}}dx_{j}+X^{i}%
\frac{\partial ^{2}L}{\partial x_{j}\partial x_{n+i}}\delta
_{i}^{j}dx_{2n+i}-X^{2n+i}\frac{\partial ^{2}L}{\partial x_{j}\partial
x_{n+i}}dx_{j}+X^{i}\frac{\partial ^{2}L}{\partial x_{j}\partial x_{i}}%
\delta _{i}^{j}dx_{3n+i} \\ 
-X^{3n+i}\frac{\partial ^{2}L}{\partial x_{j}\partial x_{i}}dx_{j}-X^{n+i}%
\frac{\partial ^{2}L}{\partial x_{n+j}\partial x_{3n+i}}\delta
_{n+i}^{n+j}dx_{i}+X^{i}\frac{\partial ^{2}L}{\partial x_{n+j}\partial
x_{3n+i}}dx_{n+j} \\ 
-X^{n+i}\frac{\partial ^{2}L}{\partial x_{n+j}\partial x_{2n+i}}\delta
_{n+i}^{n+j}dx_{n+i}+X^{n+i}\frac{\partial ^{2}L}{\partial x_{n+j}\partial
x_{2n+i}}dx_{n+j}+X^{n+i}\frac{\partial ^{2}L}{\partial x_{n+j}\partial
x_{n+i}}\delta _{n+i}^{n+j}dx_{2n+i} \\ 
-X^{2n+i}\frac{\partial ^{2}L}{\partial x_{n+j}\partial x_{n+i}}%
dx_{n+j}+X^{n+i}\frac{\partial ^{2}L}{\partial x_{n+j}\partial x_{i}}\delta
_{n+i}^{n+j}dx_{3n+i}-X^{3n+i}\frac{\partial ^{2}L}{\partial x_{n+j}\partial
x_{i}}dx_{n+j} \\ 
-X^{2n+i}\frac{\partial ^{2}L}{\partial x_{2n+j}\partial x_{3n+i}}\delta
_{2n+i}^{2n+j}dx_{i}+X^{i}\frac{\partial ^{2}L}{\partial x_{2n+j}\partial
x_{3n+i}}dx_{2n+j}-X^{2n+i}\frac{\partial ^{2}L}{\partial x_{2n+j}\partial
x_{2n+i}}\delta _{2n+i}^{2n+j}dx_{n+i} \\ 
+X^{n+i}\frac{\partial ^{2}L}{\partial x_{2n+j}\partial x_{2n+i}}%
dx_{2n+j}+X^{2n+i}\frac{\partial ^{2}L}{\partial x_{2n+j}\partial x_{n+i}}%
\delta _{2n+i}^{2n+j}dx_{2n+i}-X^{2n+i}\frac{\partial ^{2}L}{\partial
x_{2n+j}\partial x_{n+i}}dx_{2n+j} \\ 
+X^{2n+i}\frac{\partial ^{2}L}{\partial x_{2n+j}\partial x_{i}}\delta
_{2n+i}^{2n+j}dx_{3n+i}-X^{3n+i}\frac{\partial ^{2}L}{\partial
x_{2n+j}\partial x_{i}}dx_{2n+j}-X^{3n+i}\frac{\partial ^{2}L}{\partial
x_{3n+j}\partial x_{3n+i}}\delta _{3n+i}^{3n+j}dx_{i} \\ 
+X^{i}\frac{\partial ^{2}L}{\partial x_{3n+j}\partial x_{3n+i}}%
dx_{3n+j}-X^{3n+i}\frac{\partial ^{2}L}{\partial x_{3n+j}\partial x_{2n+i}}%
\delta _{3n+i}^{3n+j}dx_{n+i}+X^{n+i}\frac{\partial ^{2}L}{\partial
x_{3n+j}\partial x_{2n+i}}dx_{3n+j} \\ 
+X^{3n+i}\frac{\partial ^{2}L}{\partial x_{3n+j}\partial x_{n+i}}\delta
_{3n+i}^{3n+j}dx_{2n+i}-X^{2n+i}\frac{\partial ^{2}L}{\partial
x_{3n+j}\partial x_{n+i}}dx_{3n+j}+X^{3n+i}\frac{\partial ^{2}L}{\partial
x_{3n+j}\partial x_{i}}\delta _{3n+i}^{3n+j}dx_{3n+i} \\ 
-X^{3n+i}\frac{\partial ^{2}L}{\partial x_{3n+j}\partial x_{i}}dx_{3n+j}%
\end{array}
\label{3.30}
\end{equation}%
Since the closed quaternion K\"{a}hler form $\Phi _{L}^{H}$ on $M$ is the
symplectic structure, it is found

\begin{equation}
\begin{array}{c}
E_{L}^{H}=V_{H}(L)-L=X^{i}\frac{\partial L}{\partial x_{3n+i}}+X^{n+i}\frac{%
\partial L}{\partial x_{2n+i}}-X^{2n+i}\frac{\partial L}{\partial x_{n+i}}%
-X^{3n+i}\frac{\partial L}{\partial x_{i}}-L.%
\end{array}
\label{3.31}
\end{equation}

Hence we have

\begin{equation}
\begin{array}{c}
dE_{L}^{H}=X^{i}\frac{\partial ^{2}L}{\partial x_{j}\partial x_{3n+i}}%
dx_{j}+X^{n+i}\frac{\partial ^{2}L}{\partial x_{j}\partial x_{2n+i}}%
dx_{j}-X^{2n+i}\frac{\partial ^{2}L}{\partial x_{j}\partial x_{n+i}}%
dx_{j}-X^{3n+i}\frac{\partial ^{2}L}{\partial x_{j}\partial x_{i}}dx_{j} \\ 
X^{i}\frac{\partial ^{2}L}{\partial x_{n+j}\partial x_{3n+i}}dx_{n+j}+X^{n+i}%
\frac{\partial ^{2}L}{\partial x_{n+j}\partial x_{2n+i}}dx_{n+j}-X^{2n+i}%
\frac{\partial ^{2}L}{\partial x_{n+j}\partial x_{n+i}}dx_{n+j}-X^{3n+i}%
\frac{\partial ^{2}L}{\partial x_{n+j}\partial x_{i}}dx_{n+j} \\ 
X^{i}\frac{\partial ^{2}L}{\partial x_{2n+j}\partial x_{3n+i}}%
dx_{2n+j}+X^{n+i}\frac{\partial ^{2}L}{\partial x_{2n+j}\partial x_{2n+i}}%
dx_{2n+j}-X^{2n+i}\frac{\partial ^{2}L}{\partial x_{2n+j}\partial x_{n+i}}%
dx_{2n+j}-X^{3n+i}\frac{\partial ^{2}L}{\partial x_{2n+j}\partial x_{i}}%
dx_{2n+j} \\ 
X^{i}\frac{\partial ^{2}L}{\partial x_{3n+j}\partial x_{3n+i}}%
dx_{3n+j}+X^{n+i}\frac{\partial ^{2}L}{\partial x_{3n+j}\partial x_{2n+i}}%
dx_{3n+j}-X^{2n+i}\frac{\partial ^{2}L}{\partial x_{3n+j}\partial x_{n+i}}%
dx_{3n+j}-X^{3n+i}\frac{\partial ^{2}L}{\partial x_{3n+j}\partial x_{i}}%
dx_{3n+j} \\ 
-\frac{\partial L}{\partial x_{j}}dx_{j}-\frac{\partial L}{\partial x_{n+j}}%
dx_{n+j}-\frac{\partial L}{\partial x_{2n+j}}dx_{2n+j}-\frac{\partial L}{%
\partial x_{3n+j}}dx_{3n+j}.%
\end{array}
\label{3.32}
\end{equation}%
Using \textbf{Eq.} (\ref{1.1}), we calculate\ the following expression:

\begin{equation}
\begin{array}{c}
-X^{i}\frac{\partial ^{2}L}{\partial x_{j}\partial x_{3n+i}}dx_{j}-X^{i}%
\frac{\partial ^{2}L}{\partial x_{j}\partial x_{2n+i}}dx_{n+j}+X^{i}\frac{%
\partial ^{2}L}{\partial x_{j}\partial x_{n+i}}dx_{2n+j}+X^{i}\frac{\partial
^{2}L}{\partial x_{j}\partial x_{i}}dx_{3n+j} \\ 
-X^{n+i}\frac{\partial ^{2}L}{\partial x_{n+j}\partial x_{3n+i}}%
dx_{j}-X^{n+i}\frac{\partial ^{2}L}{\partial x_{n+j}\partial x_{2n+i}}%
dx_{n+j}+X^{n+i}\frac{\partial ^{2}L}{\partial x_{n+j}\partial x_{n+i}}%
dx_{2n+j} \\ 
+X^{n+i}\frac{\partial ^{2}L}{\partial x_{n+j}\partial x_{i}}%
dx_{3n+j}-X^{2n+i}\frac{\partial ^{2}L}{\partial x_{2n+j}\partial x_{3n+i}}%
dx_{j}-X^{2n+i}\frac{\partial ^{2}L}{\partial x_{2n+j}\partial x_{2n+i}}%
dx_{n+j} \\ 
+X^{2n+i}\frac{\partial ^{2}L}{\partial x_{2n+j}\partial x_{n+i}}%
dx_{2n+j}+X^{2n+i}\frac{\partial ^{2}L}{\partial x_{2n+j}\partial x_{i}}%
dx_{3n+j}-X^{3n+i}\frac{\partial ^{2}L}{\partial x_{3n+j}\partial x_{3n+i}}%
dx_{j} \\ 
-X^{3n+i}\frac{\partial ^{2}L}{\partial x_{3n+j}\partial x_{2n+i}}%
dx_{n+j}+X^{3n+i}\frac{\partial ^{2}L}{\partial x_{3n+j}\partial x_{n+i}}%
dx_{2n+j}+X^{3n+i}\frac{\partial ^{2}L}{\partial x_{3n+j}\partial x_{i}}%
dx_{3n+j} \\ 
+\frac{\partial L}{\partial x_{j}}dx_{j}+\frac{\partial L}{\partial x_{n+j}}%
dx_{n+j}+\frac{\partial L}{\partial x_{2n+j}}dx_{2n+j}+\frac{\partial L}{%
\partial x_{3n+j}}dx_{3n+j}=0.%
\end{array}
\label{3.33}
\end{equation}%
If a curve, defined by $\alpha :R\rightarrow M,$ is considered to be an
integral curve of $\xi ,$ then it holds the equation as follows: 
\begin{equation}
\begin{array}{c}
-X^{i}\frac{\partial ^{2}L}{\partial x_{j}\partial x_{3n+i}}dx_{j}-X^{n+i}%
\frac{\partial ^{2}L}{\partial x_{n+j}\partial x_{3n+i}}dx_{j}-X^{2n+i}\frac{%
\partial ^{2}L}{\partial x_{2n+j}\partial x_{3n+i}}dx_{j}-X^{3n+i}\frac{%
\partial ^{2}L}{\partial x_{3n+j}\partial x_{3n+i}}dx_{j} \\ 
-X^{i}\frac{\partial ^{2}L}{\partial x_{j}\partial x_{2n+i}}dx_{n+j}-X^{n+i}%
\frac{\partial ^{2}L}{\partial x_{n+j}\partial x_{2n+i}}dx_{n+j}-X^{2n+i}%
\frac{\partial ^{2}L}{\partial x_{2n+j}\partial x_{2n+i}}dx_{n+j}-X^{3n+i}%
\frac{\partial ^{2}L}{\partial x_{3n+j}\partial x_{2n+i}}dx_{n+j} \\ 
+X^{i}\frac{\partial ^{2}L}{\partial x_{j}\partial x_{n+i}}dx_{2n+j}+X^{n+i}%
\frac{\partial ^{2}L}{\partial x_{n+j}\partial x_{n+i}}dx_{2n+j}+X^{2n+i}%
\frac{\partial ^{2}L}{\partial x_{2n+j}\partial x_{n+i}}dx_{2n+j}+X^{3n+i}%
\frac{\partial ^{2}L}{\partial x_{3n+j}\partial x_{n+i}}dx_{2n+j} \\ 
+X^{i}\frac{\partial ^{2}L}{\partial x_{j}\partial x_{i}}dx_{3n+j}+X^{n+i}%
\frac{\partial ^{2}L}{\partial x_{n+j}\partial x_{i}}dx_{3n+j}+X^{2n+i}\frac{%
\partial ^{2}L}{\partial x_{2n+j}\partial x_{i}}dx_{3n+j}+X^{3n+i}\frac{%
\partial ^{2}L}{\partial x_{3n+j}\partial x_{i}}dx_{3n+j} \\ 
+\frac{\partial L}{\partial x_{j}}dx_{j}+\frac{\partial L}{\partial x_{n+j}}%
dx_{n+j}+\frac{\partial L}{\partial x_{2n+j}}dx_{2n+j}+\frac{\partial L}{%
\partial x_{3n+j}}dx_{3n+j}=0.%
\end{array}
\label{3.34}
\end{equation}%
or alternatively

\begin{equation}
\begin{array}{c}
-[X^{i}\frac{\partial ^{2}L}{\partial x_{j}\partial x_{3n+i}}+X^{n+i}\frac{%
\partial ^{2}L}{\partial x_{n+j}\partial x_{3n+i}}+X^{2n+i}\frac{\partial
^{2}L}{\partial x_{2n+j}\partial x_{3n+i}}+X^{3n+i}\frac{\partial ^{2}L}{%
\partial x_{3n+j}\partial x_{3n+i}}]dx_{j}+\frac{\partial L}{\partial x_{j}}%
dx_{j} \\ 
-[X^{i}\frac{\partial ^{2}L}{\partial x_{j}\partial x_{2n+i}}+X^{n+i}\frac{%
\partial ^{2}L}{\partial x_{n+j}\partial x_{2n+i}}+X^{2n+i}\frac{\partial
^{2}L}{\partial x_{2n+j}\partial x_{2n+i}}+X^{3n+i}\frac{\partial ^{2}L}{%
\partial x_{3n+j}\partial x_{2n+i}}]dx_{n+j}+\frac{\partial L}{\partial
x_{n+j}}dx_{n+j} \\ 
+[X^{i}\frac{\partial ^{2}L}{\partial x_{j}\partial x_{n+i}}+X^{n+i}\frac{%
\partial ^{2}L}{\partial x_{n+j}\partial x_{n+i}}+X^{2n+i}\frac{\partial
^{2}L}{\partial x_{2n+j}\partial x_{n+i}}+X^{3n+i}\frac{\partial ^{2}L}{%
\partial x_{3n+j}\partial x_{n+i}}]dx_{2n+j}+\frac{\partial L}{\partial
x_{2n+j}}dx_{2n+j} \\ 
+[X^{i}\frac{\partial ^{2}L}{\partial x_{j}\partial x_{i}}+X^{n+i}\frac{%
\partial ^{2}L}{\partial x_{n+j}\partial x_{i}}+X^{2n+i}\frac{\partial ^{2}L%
}{\partial x_{2n+j}\partial x_{i}}+X^{3n+i}\frac{\partial ^{2}L}{\partial
x_{3n+j}\partial x_{i}}]dx_{3n+j}+\frac{\partial L}{\partial x_{3n+j}}%
dx_{3n+j}=0%
\end{array}
\label{3.35}
\end{equation}%
Then we find the equations \ 
\begin{equation}
\begin{array}{l}
\,\,\frac{\partial }{\partial t}\left( \frac{\partial L}{\partial x_{i}}%
\right) +\frac{\partial L}{\partial x_{3n+i}}=0,\text{ }\frac{\partial }{%
\partial t}\left( \frac{\partial L}{\partial x_{n+i}}\right) +\frac{\partial
L}{\partial x_{2n+i}}=0,\, \\ 
\frac{\partial }{\partial t}\left( \frac{\partial L}{\partial x_{2n+i}}%
\right) -\frac{\partial L}{\partial x_{n+i}}=0,\text{ }\frac{\partial }{%
\partial t}\left( \frac{\partial L}{\partial x_{3n+i}}\right) -\frac{%
\partial L}{\partial x_{i}}=0.%
\end{array}
\label{3.36}
\end{equation}%
Thus the equations given in \textbf{Eq. }(\ref{3.36}) infer \textit{%
Euler-Lagrange equations} structured by means of $\Phi _{L}^{H}$ on
quaternion K\"{a}hler manifold $(M,V)$ and thus the triple $(M,\Phi
_{L}^{H},\xi )$ is said to be a \textit{mechanical system }on quaternion K%
\"{a}hler manifold $(M,V)$\textit{.}

\section{Conclusion}

From above, Lagrangian mechanics has intrinsically been described taking
into account a canonical local basis $\{F,G,H\}$ of $V$ on quaternion K\"{a}%
hler manifold $(M,V).$

The paths of semispray $\xi $ on the quaternion K\"{a}hler manifold are the
solutions Euler--Lagrange equations raised in (\ref{3.12}), (\ref{3.24}) and
(\ref{3.36}), and obtained by a canonical local basis $\{F,G,H\}$ of vector
bundle $V$ on quaternion K\"{a}hler manifold $(M,V)$. \ One can be proved
that these equations are very important to explain the rotational spatial
mechanics problems.


\begin{thebibliography}{9}
\bibitem{deleon} M. De Leon, P.R. Rodrigues, Methods of Differential
Geometry in Analytical Mechanics, North-Holland Mathematics Studies,
vol.152, Elsevier, Amsterdam, 1989.

\bibitem{tekkoyun} M. Tekkoyun, On Para-Euler--Lagrange and Para-Hamiltonian
Equations , Phys.\ Lett. A, Vol. 340, Issues 1-4, 2005, pp. 7-11

\bibitem{dan} D. Stahlke, Quaternions in Classical Mechanics, Phys 621.
http://www.stahlke.org/dan/phys-papers/quaternion-paper.pdf

\bibitem{yano} K. Yano, M. Kon, Structures on Manifolds, Series in Pure
Mathematics-Volume 3, World Scientific Publishing Co. Pte. Ltd., Singore,
1984.

\bibitem{burdujan} I. Burdujan, Clifford K\"{a}hler Manifolds, Balkan
Journal of Geometry and its Applications, Vol.13, No:2, 2008, pp.12-23
\end{thebibliography}
\end{document}